\documentstyle[12pt]{article}
\def\beq{\begin{equation}}
\def\eeq{\end{equation}}
\def\beqa{\begin{eqnarray}}
\def\eeqa{\end{eqnarray}}
\begin{document}
\begin{flushright}
VAND--TH-95--7\\
July 1995
\end{flushright}
\begin{center}
\Large
{\bf The Mass Spectrum in a Model with Calculable Dynamical
Supersymmetry Breaking}
\end{center}
\normalsize
\bigskip
\begin{center}
{\large  Tonnis A. ter Veldhuis
}
\\
{\sl Department of Physics \& Astronomy\\
Vanderbilt University\\
Nashville, TN 37235, USA}\\
\vspace{2cm}
\end{center}
\centerline{ABSTRACT}
\noindent
Models with dynamical supersymmetry breaking are interesting
because they may provide a solution to both the gauge hierarchy and the
fine-tuning problems.
However, because of strongly interacting dynamics, it is
in general impossible to analyze them quantitatively.
One of the few models with calculable dynamical supersymmetry
breaking is a model with SU(5) gauge symmetry and
two $10$'s and two $\bar 5$'s as the matter content.
We determine the ground state of this model,
find the vacuum energy, reveal the symmetry breaking pattern and
calculate the mass spectrum.
The supertrace mass relation is exploited to verify
the consistency of the calculated mass spectrum,
and an accidental degeneracy is explained.
\\
\vfill\eject
\section*{Introduction}

The Standard Model (SM), although extremely successful in
describing experimental data, is troubled with naturalness problems.
It is most likely an effective theory, valid up
to some energy scale $\Lambda$, above which a more fundamental
theory is necessary to accurately model nature.
Among physically motivated choices for $\Lambda$ are, for example,
the Grand Unification scale or the Planck scale.
However, the SM lacks an explanation for the fact that
the electroweak scale is much smaller than such a
choice of $\Lambda$. This shortcoming of the SM is commonly
referred to as the gauge hierarchy problem. Moreover,
quadratic divergencies in the Higgs boson mass necessitate
fine-tuning in each order of perturbation theory in order
to stabilize the hierarchy.
It is therefore more plausible that the SM breaks down
at a scale which is not much larger than the electroweak scale.

In models with global supersymmetry the technical fine-tuning
problem is solved by placing scalars in multiplets
with fermions.
In contrast to scalar masses, small fermion masses are
technically natural,
because, in general, chiral symmetries are gained when they
vanish.
The improved naturalness of supersymmetric models results in
an exact cancellation between
quadratic divergencies of fermion and scalar loops.

However, degenerate fermion and scalar masses are in striking
contrast with experimental observations. Therefore, supersymmetry
must be broken in realistic models.
In the Minimal Supersymmetric
Standard Model (MSSM) this breaking is achieved by the introduction
of ad hoc soft supersymmetry breaking terms \cite{GG}.
The nature of these terms is such
that they break supersymmetry explicitly, but do not generate quadratic
divergencies. In addition to breaking supersymmetry, they
are also instrumental in the spontaneous breaking of $SU(2)_W$.
However, the proliferation of parameters
severely limits the predictivity of the theory.
In addition, the MSSM does not provide a solution to the gauge
hierarchy problem.

Dynamical supersymmetry breaking may provide the
underlying mechanism that gives rise to the soft breaking terms.
At the same time, it may solve the gauge hierarchy problem,
since new scales are generated by dimensional transmutation.
In traditional models, dynamical supersymmetry breaking occurs in a
hidden sector, which is only coupled to the visible world
by gravitational interactions \cite{Nil}. Alternatively,
supersymmetry may be broken at a much lower scale, while the breaking
is communicated to the known world by
renormalizable gauge interactions \cite{DN}.

Although the mode of transmission of supersymmetry breaking from
the symmetry breaking sector to the known world is an
important issue, the study of models with dynamical
supersymmetry breaking is interesting by itself.
The no--renormalization theorem \cite{FL,GSR} states that if supersymmetry
is not broken at tree--level, it will not be broken at any order
of perturbation theory. In order to evade this
obstacle, non--perturbative physics is  an
indispensable ingredient of any theory with dynamical supersymmetry
breaking. Moreover, dynamical supersymmetry breaking can only occur in
models with a chiral matter content \cite{Wit}. Many supersymmetric
chiral gauge theories have therefore been analyzed \cite{ADS1,PT},
and although
dynamical supersymmetry breaking is suspected to occur
in several models, there are very few models in which this
can be shown explicitly.

One of these few is the
3--2--model \cite{ADS1,BPR}, which was used in reference \cite{DNS}
as the supersymmetry breaking sector to show the viability of
low energy supersymmetry breaking.
In this model non--perturbative physics
($SU(3)$ instantons) generates an effective term in the
superpotential which prevents the vacuum from occurring at the
origin of field space, while a renormalizable term keeps the vacuum from
running to infinity.
The D--flat directions of the model
are lifted by both the non--perturbative
and the renormalizable term in the superpotential.
If the coupling constant $\lambda$ for the renormalizable term is small
compared to the gauge coupling $g$, the vacuum expectation values will
occur at field strength $v$ large compared to the
scale $\Lambda$ at which the gauge interactions become strong, and
close to a D--flat direction.
Perturbative calculations are
reliable in this case, because the theory is weakly interacting.
Hence the model is dubbed \lq\lq calculable \rq\rq.
Moreover,
since there are no flat directions, the vacuum energy is non--zero, and
therefore supersymmetry is broken.
The spectrum of the model consists of heavy particles with
masses of the order $g v$, and light particles with masses
of the order $\lambda v$.

Of course, calculability does not imply physical relevance.
However, by studying several calculable models and comparing
their properties, insights may be gained into the general
structure of models with
dynamical supersymmetry breaking.

It is therefore interesting to  construct models
which feature dynamical supersymmetry breaking in a fashion
analogous with
the 3--2--model. However, the requirements of a unique
non--perturbative term in the superpotential and the existence
of D--flat directions which are completely lifted by the F--terms
prove to be very constraining.
In fact, the model we will discuss in this paper is
the only similar model known. It was proposed and qualitatively
analyzed by Affleck, Dine and Seiberg\cite{ADS2}. The purpose
of this paper is to elucidate the pattern of global symmetry
breaking in this model, and to explicitly calculate the vacuum
energy and mass--spectrum. It is assumed that
the non--perturbative dynamics is adequately described
by a non--perturbative term in the superpotential. Once
this term is included in the action,
supersymmetry appears to be  broken spontaneously.
The focus of this paper is on
the light spectrum in particular, with the hope that
our results will allow this model
to be used as the supersymmetry breaking sector of a complete
model with low energy supersymmetry breaking.

In Section one we will outline the model. Apart from the
$SU(5)$ gauge symmetry, the model is invariant under a global
$SU(2) \otimes U(1) \otimes U(1)$ symmetry.
Instantons generate a unique non--perturbative term in the
superpotential in addition to a renormalizable term.
The model has
D-flat directions which are completely lifted by F-terms.
As there
are no flat directions and the vacuum does not occur
at the origin of field space nor at infinity, supersymmetry is spontaneously
broken.
The observables in the light sector of the model are determined
in terms of only two parameters; the scale $\Lambda$ at which
the gauge interactions become strong and a Yukawa type
coupling constant $\lambda$.

In Section two the possible symmetry breaking patterns of the
model will be reviewed. The symmetry breaking pattern determines
how the spectrum is divided into a light and a heavy sector, and it
is intimately connected to the existence of various massless particles.
The results concerning  possible mass spectra will provide the
framework for the interpretation of our numerical work.
The gauge symmetry of the model is completely broken,
while the
global symmetries are broken into at most a $U(1)$ symmetry.
In order to determine which symmetry breaking pattern is actually realized,
i.e. whether or not there is a remaining $U(1)$ symmetry, it is
necessary to explicitly find the minimum of the scalar potential.

In Section three we will present our results. We numerically
minimize the scalar potential and calculate the vacuum energy.
The symmetry breaking pattern is revealed to
be $SU(2) \otimes U(1) \otimes U(1) \rightarrow U(1)$.
The masses of scalars, fermions and vector bosons are calculated,
and the light spectrum is found to include twelve scalars and six fermions.
Among the light scalars are four massless Goldstone bosons associated
with the broken generators of global symmetries. The light fermion spectrum
includes a neutral and a charged massless fermion.
The neutral massless fermion is a Goldstino, the Goldstone particle
associated
with the spontaneous breaking of supersymmetry,
whereas the charged neutral fermion saturates an 't Hooft anomaly
matching condition \cite{Hoo} for the remaining $U(1)$ symmetry.
In order to check the consistency of the
calculated mass spectrum, we explicitly verify
that it satisfies the supertrace mass relation, which is valid even
though supersymmetry is spontaneously broken \cite{FGP}.
The light spectrum contains several degeneracies in addition
to the degeneracies dictated by the remaining $U(1)$ symmetry.
These additional degeneracies are shown to be accidental.

\section{The model}

The model we study is an $SU(5)$ chiral gauge theory \cite{ADS2}.
The matter content consists of two chiral fields in
the 10 representation of $SU(5)$, and two fields in the $\bar{5}$
representation.
These matter fields will be denoted by $T_{\alpha}^{ij}$
and $\bar{F}_{i}^{\alpha}$ respectively, where the
Roman superscripts $i,j=1,..,5$ are $SU(5)$ indices and
the Greek subscript $\alpha=1,2$
is a flavor index.
Defining
\beqa
V_T & = & g V^a G_{10}^a, \nonumber \\
V_F & = & g V^a G_{5}^a,
\eeqa
with $V^a$ the twenty--four $SU(5)$ vector multiplets and
$G^a$ the generators in the appropriate representation,
the K\"{a}hler potential takes the conventional form
\beq
K= \bar{T}^{\alpha} e^{-2V_T} T_{\alpha}
   + F_{\alpha} e^{2V_F} \bar{F}^{\alpha}.
\eeq
Without loss of generality, the gauge invariant
renormalizable superpotential for this model is given by
\beq
W_{p}=\lambda \epsilon_{\alpha \beta}\bar{F}_{i}^{\alpha}
T_1^{ij} \bar{F}_{j}^{\beta}.
\eeq
A similar term with $T_2$ instead of $T_1$ can always be eliminated
by a field redefinition. Apart from the gauge symmetry, this model has
a global $SU(2)_F \otimes U(1)_A \otimes U(1)_R$ symmetry. The
$SU(2)_F$ is a flavor symmetry which transforms the $\bar{F}$ fields
into each other. Under $U(1)_A$
the fields transform as
\beqa
\bar{F}^{\alpha} & \rightarrow & e^{\frac{3}{5} \omega i} \bar{F}^{\alpha}
\nonumber \\
T_1 & \rightarrow & e^{-\frac{6}{5} \omega i } T_1
\nonumber \\
T_2 & \rightarrow & e^{\frac{4}{5} \omega i } T_2,
\eeqa
while  under the R--symmetry $U(1)_R$ the fields transform as
{\footnote{The R-weight of the gaugino is defined to be $+1$.}}
\beqa
\bar{F}^{\alpha} & \rightarrow & e^{-4 \omega i} \bar{F}^{\alpha}
(\theta e^{- \omega i}),
\nonumber \\
T_1 & \rightarrow & e^{10 \omega i }  T_1(\theta e^{- \omega i}),
\nonumber \\
T_2 & \rightarrow & e^{-8 \omega i }  T_2(\theta e^{- \omega i}).
\eeqa
Note that the renormalizable term in the superpotential explicitly
breaks an $SU(2)_T$ flavor symmetry that transforms $T_1$ into
$T_2$.

Instantons generate a unique effective non--perturbative term in the
superpotential of the form
\beq
W_{np}=\frac{\Lambda^{11}}{\Delta},
\eeq
with
\beq
\Delta=\epsilon_{\alpha \beta}\epsilon_{abcde}\epsilon_{ijklm}
\bar{F}_{r}^{\alpha} \bar{F}_{s}^{\beta} T_{1}^{ri} T_{1}^{bc} T_{1}^{de}
 T_{2}^{sa} T_{2}^{jk} T_{2}^{lm}.
\eeq
Here $\Lambda$ is the scale at which the $SU(5)$ gauge interactions
become strong and the exponent eleven gives $W_{np}$ the
correct dimension. The requirement that
$W_{np}$ is invariant under the $U(1)_R$ symmetry implies
that the power of $\Delta$ is uniquely
determined to be one. Moreover, $W_{np}$ is a singlet under $SU(2)_T$.

The theory has D--flat directions, which are lifted
by both the renormalizable and the non--perturbative terms
in the superpotential \cite{ADS2}. In the limit $\lambda \rightarrow 0$
the low energy effective theory is obtained by constraining
the fields to these D--flat directions. In this picture, the
superpotential is considered a perturbation.
The  D--flat directions of the model are solutions to the equation
\beq
T_{1ij}^{\dagger} T_{1}^{kj}
+ T_{2ij}^{\dagger} T_{2}^{kj}
- F_{1}^{\dagger k} F_{1i}
- F_{2}^{\dagger k} F_{2i} \sim \delta_{i}^{k}. \label{eq:dflat}
\eeq
Solutions to Eq.(\ref{eq:dflat}) are up to a gauge transformation
described by twelve real parameters.
The six composite objects
$J_1^{\alpha}=\epsilon_{ijklm}\bar{F}_n^{\alpha} T_1^{ij} T_1^{kl} T_2^{mn}$
,
$J_2^{\beta}=\epsilon_{ijklm}\bar{F}_n^{\beta} T_2^{ij} T_2^{kl} T_1^{mn}$
,
$X_1  =  \epsilon_{\alpha \beta} \bar{F}_i^{\alpha} T_1^{ij}
\bar{F}_j^{\beta}
$
and
$
X_2  =  \epsilon_{\alpha \beta} \bar{F}_i^{\alpha} T_2^{ij}
\bar{F}_j^{\beta}
$
provide  a convenient gauge invariant parametrization of the manifold of
D--flat directions.
The low energy theory can be described by a sigma model with
these six composite objects as coordinates.
This sigma model  has a K\"{a}hler potential
\beq
K_{eff}=K_{eff}\left( \bar{X}^{\alpha} X_{\alpha}, \bar{J}_{\beta}^{\alpha}
          J_{\alpha}^{\beta} \right),
\eeq
and a superpotential
\beq
W_{eff} = \frac{\Lambda^{11}}{\epsilon_{\alpha \beta} J_1^{\alpha}
    J_2^{\beta}} + \lambda X_1.
\eeq
The functional form of the K\"{a}hler potential $K_{eff}$ can
in principle be found by
using Eq.(\ref{eq:dflat}) to project the K\"{a}hler potential of the
full theory onto the coordinates. This is possible because in the
limit of vanishing superpotential
(no supersymmetry breaking) Eq.(\ref{eq:dflat})  is
promoted from an equation in terms of scalar components only to a superfield
equation.
However, in practice, the
complexity of the projection procedure is formidable.

Another approach to find the ground state of the theory and its
low energy properties is to minimize the scalar potential in the
D--flat directions only. This approach requires an explicit
parametrization of the D--flat directions.
Eight of the twelve parameters can be chosen
as the parameters of an $SU(2)_F \otimes SU(2)_T \otimes U(1)_A
\otimes U(1)_R$ transformation. Therefore, in order to provide
a full parametrization of the D--flat manifold, a solution to
Eq.(\ref{eq:dflat}) with four parameters is required.
A specific
example of a flat direction with two parameters is \cite{ADS2}
\begin{eqnarray}
T_{1}=\left(
\begin{array}{ccccc}
0 & a & 0 & 0 & 0 \\
-a & 0 & 0 & 0 & 0 \\
0 & 0 & 0 & c & 0 \\
0 & 0 & -c & 0 & 0 \\
0 & 0 & 0 & 0 & 0
\end{array}
\right),
T_{2}=\left(
\begin{array}{ccccc}
0 & 0 & 0 & 0 & c \\
0 & 0 & 0 & b & 0 \\
0 & 0 & 0 & 0 & 0 \\
0 & -b & 0 & 0 & 0 \\
-c & 0 & 0 & 0 & 0
\end{array}
\right),
F_{1}=
\left(
\begin{array}{c}
a \\
0 \\
0 \\
0 \\
0
\end{array}
\right),
F_{2}=
\left(
\begin{array}{c}
0 \\
0 \\
0 \\
b \\
0
\end{array}
\right),
\end{eqnarray}
with $c=\sqrt{a^2+b^2}$.
Unfortunately, we
were unable to determine a solution with four parameters.
%Evaluating the scalar potential in this
%direction and minimizing with respect to the parameters $a$ and $b$ yields
%\beq
%a=b=\frac{1}{2} \Lambda (\frac{6}{\lambda})^{\frac{1}{11}},
%\eeq
%and
%\beq
%V=\frac{11}{18} 6^{\frac{4}{11}} \Lambda^4 \lambda^{\frac{18}{11}}.
%\eeq
Proceeding regardless and minimizing the potential with respect to the
parameters
$a$ and $b$ shows
that the D--flat directions are lifted
, that the vacuum expectation values of the
fields scale as $v \sim \lambda^{-\frac{1}{11}} \Lambda$ and
that the vacuum energy scales as $V \sim \lambda^2 v^4$.

However, in the absence of  a general parametrization of the
D--flat directions,
we determine the low energy
properties of the model by numerically minimizing the
full scalar potential, including both D and F terms, with respect
to all scalar fields.

\section{Qualitative analysis of the mass spectrum}

Many properties of the structure of the mass--spectrum are
determined by the symmetry breaking pattern \cite{ADS2}, although some
aspects require an explicit minimization of the scalar potential.
It is important to observe that the gauge symmetry is completely
broken \cite{ADS2}. This follows from the fact that the quantity $\Delta$
vanishes
at points in field space where the gauge symmetry is only partially broken.
The scalar potential contains terms inversely proportional to
$\Delta$, and therefore blows up at points with residual gauge
symmetry.

In the supersymmetric limit, that is in the absence
of a superpotential, the Higgs
mechanism  causes the component fields to rearrange into
%A massless vector multiplet
%contains one fermion and one massless vector. A massive vector
%multiplet in contrast contains one scalar,
%two fermions and a massive vector.
twenty-four massive vector multiplets with masses of the
order $g v$ and six massless
chiral multiplets.

%Now we have to do some counting.
%Each chiral superfield in the $10$ representation
%contains 10 complex scalars,
%and each chiral field in the $\bar{5}$ representation contains
%5 complex scalars, therefore there are a total of 60 real scalars.
%Adding the number of fermions in the chiral superfields to the
%24 gauginos yields a total of 54 fermions. According to the
%Higgs mechanism, 24 scalars will be "eaten" to give mass to the
%vector bosons, and 24 more scalars will be heavy, with masses
%degenerate with the vector bosons, and the remaining 12 scalars
%are massless. Of the 54 fermions, 48 will be heavy with masses
%degenerate with the vector bosons, and the remaining 6 will be massless.

Of course, when the superpotential is switched on, supersymmetry
is broken.
In particular, some of the twelve scalars and six fermions that are
massless in the supersymmetric limit obtain masses proportional
to $\lambda v$.
The number of modes that remain massless depends on which global symmetries
are broken. In order to study
this issue, it is useful
to observe that $\Delta=\epsilon_{\alpha \beta} J_1^{\alpha}J_2^{\beta}$.
The quantities $J_1^{\alpha}$ and $J_2^{\alpha}$
transform as a doublet under $SU(2)_F$.
As $\Delta$ is unequal to zero at the minimum, at least
$(J_1^1,J_2^2)$ or $(J_1^2,J_2^1)$ is unequal to zero.
If either $(J_1^1,J_2^2)$ or $(J_1^2,J_2^1)$ is equal to zero\footnote{
Actually, in general an $SU(2)_F$ transformation is
needed to bring $J_{\alpha}^{\beta}$ into this form.}
then the global symmetries are broken into a remaining
$Q=I_Z \pm \frac{A}{2}$; if both $(J_1^1,J_2^2)$ and $(J_1^2,J_2^1)$
are not equal to zero, then the global symmetries are completely
broken.
As a consequence there are either four or five Goldstone bosons.

The anomaly of the $U(1)_Q$ symmetry in the fundamental theory is
$\sum Q^3 = \pm 1$. If this symmetry is not broken, then
this anomaly needs to be matched in the effective
low--energy theory \cite{Hoo}.
Therefore, if there is a remaining $U(1)_Q$ symmetry, then
the spectrum contains a massless fermion with (negative) unit charge.
In addition,
the spectrum contains a massless neutral fermion associated with
the spontaneous breaking of global supersymmetry, the Goldstino.

\section{The particle spectrum}

We numerically minimized the scalar potential. In the limit
$\lambda<<g$, the scalar potential has very narrow valleys, which
results in slow convergence of the minimization procedure.
We therefore chose to first minimize the potential for
values of $\lambda$ and $g$ not too far apart. Using this
location of the  minimum as an initial condition,
we then increased the value of $g$ and minimized the
potential again. We repeated this procedure until
the location and value of the minimum did not change significantly
if $g$ was raised further.
The vacuum energy was found to be $2.806 \Lambda^4 \lambda^{\frac{18}{11}}$.
In order to determine the global symmetry breaking pattern, we calculated the
vacuum expectation values of the composite structures
$J_1^{\alpha}$, $J_2^{\alpha}$,
$X_1 $
and
$
X_2 $.
%J_1^{\alpha} & = &
%\epsilon_{abcde} \bar{F}_r^{\alpha} T_1^{bc} T_1^{de} T_2^{sa} \nonumber\\
%J_2^{\alpha} & = &
%\epsilon_{abcde} \bar{F}_r^{\alpha} T_2^{bc} T_2^{de} T_1^{sa}. \nonumber
%\eeqa
These expectation values are listed in Table \ref{tab:vevs} together
with the quantum numbers of the corresponding objects.
It is clear from Table \ref{tab:vevs} that the vacuum expectation values
of objects which transform
non-trivially under $Q=I_Z - \frac{A}{2}$
vanish. The global symmetries are therefore broken into a
single remaining $U(1)_Q$. Although Table \ref{tab:vevs}
gives the results in the limit $\frac{\lambda}{g} \rightarrow 0$,
$U(1)_Q$ is also a good symmetry for finite ratios of $\frac{\lambda}{g}$.
In contrast, the vacuum expectation values of $J_1^2$ and $J_2^1$ are
only equal in the
limit $\frac{\lambda}{g} \rightarrow 0$. This degeneracy is not
dictated by the $U(1)_Q$ symmetry.
In order to explain this degeneracy, it is useful to
note that the object $J_{\alpha}^{\beta}$ transforms as
a doublet under both
$SU(2)_T$ and $SU(2)_F$.
The vacuum expectation value of $J_{\alpha}^{\beta}$ breaks the symmetry
group $SU(2)_T \otimes SU(2)_F$
into a diagonal subgroup $SU(2)_D$. Under this subgroup,
$( J_1^1, J_1^2-J_2^1, J_2^2 )$ transforms as a triplet, while
$J_1^2+J_2^1$ is a singlet. Although $SU(2)_T$ is explicitly broken
by the renormalizable term in the superpotential,
this does not feed into the expectation values of $J_{\alpha}^{\beta}$
at tree level. As a consequence, the light mass-spectrum contains some
accidental degeneracies.
\begin{table}
\begin{center}
\begin{tabular}{|l|l|l|l|l|} \hline
       & $A$   & $I_Z$   & $Q=I_Z-\frac{A}{2}$  & vev  \\ \hline
$X_1$  & 0     & 0       & 0          & $0.124 \Lambda^3 \lambda^{-3/11}$\\
$X_2$  & 2     & 0       & -1         & 0 \\
$J_1^1$& -1    & 1/2     & 1          & 0\\
$J_1^2$& -1    & -1/2    & 0          & $-2.301 \Lambda^4 \lambda^{-4/11}$\\
$J_2^1$& 1     & 1/2     & 0          & $-2.301 \Lambda^4 \lambda^{-4/11}$ \\
$J_2^2$& 1     & -1/2    & -1         & 0\\
\hline
\end{tabular}
\caption{Vacuum expectation values and quantum numbers
of some composite structures.
\label{tab:vevs} }
\end{center}
\end{table}

We next calculated the scalar mass matrix by numerically evaluating
%$\frac{\partial^2 V}{\partial \phi_i \partial \phi_j}$
the second derivative of the scalar potential with
respect to all sixty real scalars at the
minimum.
After diagonalizing this matrix we found twenty-four masses of the order
$g v$, eight masses of the order $\lambda v$ and twenty-eight
masses equal to zero.
Twenty-four of the twenty-eight massless scalars are
would be Goldstone bosons which are eaten by the
$SU(5)$ vector bosons. The spectrum contains therefore
four genuine Goldstone bosons,
in accordance with the symmetry breaking pattern.
One pair of these Goldstone bosons has a charge under
$U(1)_Q$. The remaining eight light scalars include two charged pairs.
The light scalar masses are listed  in
the limit $\frac{\lambda}{g} \rightarrow 0$ in Table \ref{tab:sca}.
The degeneracy of the masses of one charged pair of scalars and a
neutral scalar is accidental. The corresponding states form a
triplet under $SU(2)_D$, and the degeneracy is lifted for finite
values of $\frac{\lambda}{g}$.

\begin{table}
\begin{center}
\begin{tabular}{|l|l|l|} \hline
    & mass  & Q  \\ \hline
1,2 & 0     & $\pm 1$ \\
3   & 0     & 0 \\
4   & 0     & 0 \\
5,6 & 2.55  & $\pm 1$ \\
7   & 2.74  & 0 \\
8,9 & 2.74  & $\pm 1$ \\
10  & 3.90  & 0 \\
11  & 5.95  & 0 \\
12  & 9.32  & 0 \\ \hline
\end{tabular}
\caption{Masses, in units of $\lambda^{10/11} \Lambda$,
and charges of the light scalars. \label{tab:sca}}
\end{center}
\end{table}

The fermion mass terms are of the form
$\frac{1}{2} \frac{\partial^2 W}{\partial \phi_i \partial \phi_j} \psi_i
\psi_j$
and $i g \phi^{\dagger i} G_i^{aj} \psi_j \lambda^a$,
where $\psi_j$ are the thirty matter fermions and $\lambda^a$ are
the twenty-four $SU(5)$ gauginos.
We diagonalized the fermion mass matrix,
and found forty-eight fermion masses of the
order $g v$, four fermion masses of the order $\lambda v$
and two fermion masses equal to zero.
One of the massless fermions is neutral, and can be identified as
the Goldstino. The other massless fermion is charged and saturates
the 't Hooft anomaly matching condition.
Two
of the four remaining light fermions are charged.
The light fermion spectrum in the limit $\frac{\lambda}{g} \rightarrow 0$
is summarized
in Table \ref{tab:fer}.
As in the case of the scalars, the degeneracy of the masses of a pair
of charged fermions and a neutral fermion is accidental

\begin{table}
\begin{center}
\begin{tabular}{|l|l|l|} \hline
    & mass & Q  \\ \hline
1   & 0      & 0 \\
2   & 0      & $-1$ \\
3,4 & 0.716  & $\pm 1$ \\
5   & 0.716  & 0 \\
6   & 7.486  & 0 \\ \hline
\end{tabular}
\caption{Masses, in units of $\lambda^{10/11} \Lambda$,
and charges of the light fermions. \label{tab:fer}}
\end{center}
\end{table}

As a consistency check of our calculation
we calculated the vector boson
masses and verified that the spectrum satisfies
the supertrace mass relation \cite{FGP}
\beq
\sum m_{scalar}^2 + 3\sum m_{vector}^2 - 2 \sum m_{fermion}^2=0.
\eeq
Here the sums have to be taken over both the light and the
heavy particles.
%$V^a V^b G_l^{ai} G_i^{bk} \phi_j \phi^{\dagger l}$.
As an additional check we minimized the scalar potential
in the absence of F-terms.
In this case supersymmetry is not
broken and we found a mass spectrum
consistent with twenty-four massive
vector multiplets and six massless chiral multiplets,
as expected.

To conclude, we discuss the various massless states in the
low energy theory.
In the framework of a theory with local supersymmetry,
the R--symmetry is necessarily broken, and the R--axion
will obtain a mass \cite{BPR}. At the same time,
the Goldstino
will be eaten by the gravitino.
The remaining Goldstone bosons will be eaten by vector bosons
if the corresponding symmetries are gauged.
Finally, the charged massless fermion will disappear from the
spectrum if the $U(1)_Q$ symmetry is gauged and appropriate matter
is added to cancel its anomaly.

The next challenge is to construct a realistic visible sector model
using this model as the symmetry breaking sector!

\section*{Acknowledgements}
The author thanks Bill Bardeen, Tom Clark, Tom Kephart, Tom Weiler
and Stuart Wick
for useful discussions. The hospitality of
Fermilab and the Aspen Center for Physics, where part of this work
was done, is gratefully acknowledged.
This work was supported in part by the U.S. Department
of Energy under grant No. DE-FG05-85ER40226.


\begin{thebibliography}{99}
\bibitem{GG}  L. Girardello and M.T. Grisaru
              Nucl. Phys. {\bf B194} (1982) 65.
\bibitem{Nil} H. Nilles, Phys. Rep. {\bf 110} (1984) 1.
\bibitem{DN}  M. Dine and A.E. Nelson, Phys. Rev. {\bf D47} (1993) 1277.
\bibitem{FL}  K. Fujikawa and W. Lang, Nucl. Phys. {\bf B88} (1975) 61.
\bibitem{GSR} M.J. Grisaru, W. Siegel and M. Roc\v{o}k,
              Nucl. Phys. {\bf B159} (1979) 429.
\bibitem{Wit} E. Witten, Nucl. Phys. {\bf212} (1982) 253.
\bibitem{ADS1} I. Affleck, M. Dine and N. Seiberg, Nucl. Phys. {\bf B256}
               (1985) 557.
\bibitem{PT}   E. Poppitz and S.P. Trivedi, EFI--95--44,
               Fermilab--Pub--95/258--T.
%general study of DSB
%\bibitem{DSS}  S. Dimopoulos, S. Raby and L. Susskind, Nucl. Phys {\bf B173}
%               (1980) 208.
%complementarity, the conjectured equivalence between Higgs and confined
%phases
\bibitem{BPR}  J. Bagger, E. Poppitz and L. Randall,
               Nucl.Phys. {\bf B426} (1994) 3.
%the axion problem, discussion of Su(3)xSu(2) model.
\bibitem{DNS}  M. Dine, A.E. Nelson and Y. Shirman,
               Phys.Rev. {\bf D51} (1995) 1362.
\bibitem{ADS2} I. Affleck, M. Dine and N. Seiberg, Phys. Rev. Lett. {\bf 52}
               (1984) 1677.
%extensive qualitative analysis of the SU(5) model
%model of low energy visible sector dynamical susy breaking; SU(3)xSU(2)
\bibitem{Hoo}  G. 't Hooft, In Recent Developments in Gauge Theories,
               edited by G. 't Hooft et al. (Plenum, New York, 1980).
\bibitem{FGP}  S. Ferrara, L. Girardello and F. Palumbo,
               Phys. Rev. {\bf D20} (1979) 403.
\end{thebibliography}
\end{document}